\documentstyle[epsfig,epsf,psfig,aps,preprint]{revtex}
\newcommand{\bea}{\begin{eqnarray}}
\newcommand{\eea}{\end{eqnarray}}
\newcommand{\be}{\begin{equation}}
\newcommand{\ee}{\end{equation}}
\newcommand{\bt}{\begin{tabular}}
\newcommand{\et}{\end{tabular}}
\newcommand{\Tr}{{\rm Tr}}
\newcommand{\no}{\nonumber}
\newcommand{\ovl}{\overline}

\newcommand{\pa}{\partial}
\newcommand{\beas}{\begin{eqnarray*}}
\newcommand{\eeas}{\end{eqnarray*}}

\begin{document}
\title{Hadrons in Dense Resonance-Matter: A Chiral $SU(3)$ Approach}
\author{D. Zschiesche$^1$, P. Papazoglou$^1$, 
S. Schramm$^1$, J. Schaffner-Bielich$^2$, H.~St\"ocker$^1$ and W.~Greiner$^1$}
  \address{
        $^1$Institut f\"ur Theoretische Physik, 
        $^{~}$Postfach 11 19 32, D-60054 Frankfurt am Main, Germany\\
        $^2$Riken BNL Research Center, Brookhaven National Lab, Upton,
New York 11973} 
\maketitle
\begin{abstract}
A nonlinear chiral SU(3) approach including the spin $\frac 32$
decuplet is developed to describe dense matter. The coupling
constants of the baryon resonances to the scalar mesons are determined
from the decuplet vacuum masses and SU(3) symmetry relations. Different
methods of mass generation show significant differences in the
properties of the spin-$\frac 32$ particles and in the nuclear equation of 
state.
\end{abstract}
\draft
\section{Introduction}
The investigation of the equation of state of strongly interacting matter 
is one of the most challenging problems in nuclear and heavy ion physics. 
Dense nuclear matter exists in the 
interior of neutron stars, and its behaviour plays a crucial role for the 
structure and properties of these stellar objects.  The
behaviour of hadronic matter at high densities and temperatures strongly 
influences the observables in relativistic
heavy ion collisions (e.g. flow, particle production,...). 
The latter depend on the bulk and nonequilibrium properties of the
produced matter (e.g. pressure, density, temperature, viscosity,...) and the 
properties of the constituents (effective masses, decay widths,
dispersion relations,...).
So far it is not possible to determine the equation of state of hadronic 
matter at high densities (and temperatures) from first principles. 
QCD is not solvable in the regime of low momentum transfers and finite
baryon densities. Therefore one has to pursue alternative ways to
describe the hadrons in dense matter. Effective
models, where only the relevant degrees of freedom for the problem are
considered are solvable and can
contain the essential characteristics of the full theory. For the
case of strongly interacting matter this means that one considers 
hadrons rather than quarks and gluons as the relevant degrees of
freedom.
Several such models like the RMF model(QHD) and its extensions
(QHD II, nonlinear Walecka model) successfully describe nuclear matter and
finite nuclei \cite{wale74,sero86,boguta77,bogustoe,fpw}. 
Although these models are effective relativistic 
quantum field theories of baryons and mesons, they do not consider 
essential features of QCD, namely broken scale invariance and 
approximate chiral symmetry. 
Including SU(2) chiral symmetry in these models
by adding repulsive vector mesons to the $SU(2)$-linear
$\sigma$-model does neither lead to a reasonable description of nuclear
matter ground state properties nor of finite nuclei \cite{kerm74}. 
Either one must use a nonlinear realization of chiral symmetry 
\cite{furn93,tang95} or include a dilaton field and a logarithmic
potential motivated by broken scale invariance \cite{sche80,heid94} in
order 
to obtain a satisfactory description of nuclear matter. 
Extending these approaches to the strangeness sector leads to a number of
new, undetermined coupling constants due to the additional strange hadrons.
Both to overcome this problem and to put restrictions on
the coupling constants in the non-strange sector the inclusion of SU(3)
\cite{scha93} and chiral SU(3) \cite{paper2,paper3} has been
investigated in the last years. 
Recently \cite{paper3} it was shown that an extended 
$SU(3) \times SU(3)$ chiral $\sigma-\omega$ model can
describe nuclear matter ground state properties, vacuum properties and
finite nuclei simultaneously. This model includes the lowest lying
SU(3) multiplets of the baryons (octet), the spin-0 and the spin-1 mesons
(nonets) as physical degrees of freedom. The present paper will
discuss the predictions of this model for high density nuclear matter,
including the spin $\frac 32$ baryon resonances
(decuplet). This is necessary, 
because the increasing nucleonic fermi levels make the production of 
resonances energetically favorable at high densities. 
The paper is structured as follows: Section II summarizes
the nonlinear chiral $SU(3) \times SU(3)$-model. Section III
gives the baryon meson interaction, with main focus on the  baryon
meson-decuplet interaction and the constraints on the additional
coupling constants. In section IV the resulting
equations of motions and thermodynamic observables in the mean field
approximation are discussed. Section V contains the results for dense
hadronic matter, followed by the conclusions.
\section{Lagrangian of the nonlinear chiral SU(3) model}
\label{lagrangian}
We use a relativistic field theoretical model of baryons and mesons based on
chiral symmetry and scale invariance to describe strongly interacting
nuclear matter. In earlier work the Lagrangian including the baryon octet,
the spin-0 and spin-1 mesons has been developed \cite{paper3}. Here the
additional inclusion of the spin-$\frac{3}{2}$ baryon decuplet 
 for infinite nuclear matter will be discussed. The general form of the
Lagrangian then looks as follows:
\be
\label{lagrange}
{\cal L} = {\cal L}_{\mathrm{kin}}+\sum_{W=X,Y,V,{\cal A},u}{\cal L}_{\mathrm{BW}}+
{\cal L}_{\mathrm{VP}}
+{\cal L}_{\mathrm{vec}}+{\cal L}_{0}+{\cal L}_{\mathrm{SB}} .\no
\ee
${\cal L}_{\mathrm{kin}}$ is 
the kinetic energy term, ${\cal L}_{\mathrm{BW}}$ includes the  
interaction terms of the different baryons with the various spin-0 and spin-1 
mesons. ${\cal L}_{\rm{VP}}$ contains the interaction terms 
of vector mesons with pseudoscalar mesons. 
${\cal L}_{\rm{vec}}$ generates the masses of the spin-1 mesons through 
interactions with spin-0 mesons, and ${\cal L}_{0}$ gives the meson-meson 
interaction terms which induce the spontaneous breaking of chiral symmetry.
It also includes the scale breaking logarithmic potential. Finally, 
${\cal L}_{\mathrm{SB}}$ introduces an explicit symmetry breaking of the
U(1)$_A$ symmetry, the SU(3)$_V$ symmetry, and the chiral symmetry. 
These terms have been discussed in detail in \cite{paper3} and this
shall not be repeated here. We will concentrate on the new terms in 
${\cal L}_{\mathrm{BW}}$, which are due to adding the
baryon resonances. 
\section{Baryon meson interaction}
${\cal L}_{BW}$ consists of the interaction terms of the included baryons
(octet and decuplet) and the mesons (spin-0 and spin-1).
For the spin-$\frac 12$ baryons the $SU(3)$ structure of the 
couplings to all mesons are the same, except for the
difference in Lorentz space. 
For a general meson field $W$ they read
\be
\label{spin12meson}
{\cal L}_{\mbox{OW}} = 
-\sqrt{2}g_{O8}^W \left(\alpha_{OW}[\ovl{B}{\cal O}BW]_F+ (1-\alpha_{OW}) 
[\ovl{B} {\cal O}B W]_D \right)
- g_{O1}^W \frac{1}{\sqrt{3}} \Tr(\ovl{B}{\cal O} B)\Tr W  \, ,  
\ee
with $[\ovl{B}{\cal O}BW]_F:=\Tr(\ovl{B}{\cal O}WB-\ovl{B}{\cal O}BW)$ and 
$[\ovl{B}{\cal O}BW]_D:= \Tr(\ovl{B}{\cal O}WB+\ovl{B}{\cal O}BW) - \frac{2}{3} 
\Tr (\ovl{B}{\cal O} B) \Tr W$.
The different terms to be considered are those for the interaction
of spin-$\frac 12$ baryons ($B$),  with
scalar mesons ($W=X, {\cal O}=1$), with 
vector mesons  ($W=V_{\mu}, {\cal O}=\gamma_{\mu}$),
with axial vector mesons ($W={\cal A}_\mu, {\cal O}=\gamma_\mu \gamma_5$)
and with
pseudoscalar mesons ($W=u_{\mu},{\cal O}=\gamma_{\mu}\gamma_5$), respectively.
For the spin-$\frac 32$ baryons ($D^\mu$) one can construct a coupling term
similar to (\ref{spin12meson}) 
\be
\label{spin32meson}
{\cal L}_{\mbox{DW}} = 
-\sqrt{2}g_{D8}^{W} [\ovl{D^\mu}{\cal O}D_\mu W] 
- g_{D1}^W [\ovl{D^\mu}{\cal O}D_\mu ] Tr W \, ,
\ee
where $[\ovl{D^\mu}{\cal O}D_\mu W]$ and $[\ovl{D^\mu}{\cal O}D_\mu]$
are obtained from coupling 
$[\bar{10}]\times[10]\times[8]=[1]+[8]+[27]+[64]$ 
and $[\bar{10}]\times[10]\times[1]$  
to an SU(3) singlet, respectively.
In the following we focus on the
couplings of the baryons to the scalar mesons which dynamically generate the
hadron masses and vector mesons
which effectively describe the short-range repulsion.
For the pseudoscalar
mesons only a pseudovector coupling is possible, since in the
nonlinear realization of chiral symmetry \cite{paper3} they only 
appear in derivative terms. Pseudoscalar and axial mesons
have a vanishing expectation value at the mean field level, so that their
coupling terms will not be discussed in detail here.

\label{bm}
\subsection*{Scalar Mesons}
The baryons and the scalar mesons 
transform equally in the left and right subspace.  
Therefore, in contrast to the linear realization of chiral symmetry, 
an $f$-type coupling is allowed for the baryon-octet-meson interaction. 
In addition, it is possible to construct mass terms for 
baryons and to couple them to chiral singlets. Since the current quark
masses in QCD are small compared to the hadron masses, we will use
baryonic mass terms only as small corrections to the dynamically
generated masses. Furthermore a coupling of the
baryons to the dilaton field $\chi$ is also possible, but this will 
be discussed in a later publication.
After insertion of the vacuum matrix 
$\langle X\rangle$, (Eq.\ref{vev}), one obtains the baryon masses as 
generated by the vacuum expectation value (VEV) of the two meson fields:
\bea
\label{bmver1eq1}
 m_N        &=& m_0 -\frac{1}{3}g_{O8}^S(4\alpha_{OS}-1)(\sqrt{2}\zeta-\sigma) \\ \no
 m_{\Lambda}&=& m_0-\frac{2}{3}g_{O8}^S(\alpha_{OS}-1)(\sqrt{2}\zeta-\sigma) \\ \no
 m_{\Sigma} &=& m_0+\frac{2}{3}g_{O8}^S(\alpha_{OS}-1)(\sqrt{2}\zeta-\sigma)  \\ \no
 m_{\Xi}    &=& m_0+\frac{1}{3}g_{O8}^S(2\alpha_{OS}+1)(\sqrt{2} \zeta-\sigma) \no
\eea
with $m_0=g_{O1}^S(\sqrt{2} \sigma+\zeta)/\sqrt{3}$.
The parameters $g_{O1}^S$, $g_{O8}^S$ and $\alpha_{OS}$ can be used to fit the 
baryon-octet masses 
to their experimental values. Besides the current quark mass terms
discussed in \cite{paper3}, no additional explicit symmetry 
breaking term is needed. Note that the nucleon mass depends on the 
{\it strange condensate $\zeta$!} For $\zeta=\sigma/\sqrt{2}$ (i.e. 
$f_{\pi}=f_K$), the masses are degenerate, and the vacuum is 
SU(3)$_V$-invariant. For the spin-$\frac 32$ baryons the procedure is similar.
If the vacuum matrix for the scalar condensates is inserted one
obtains the dynamically generated vacuum masses of the baryon decuplet 
\bea
\label{resmassen}
 m_{\Delta} &=& g_D^S \left[(3-\alpha_{DS})\sigma+\alpha_{DS}
\sqrt{2}\zeta\right]\\ \no
 m_{\Sigma^\ast}&=& g_D^S \left[ 2\sigma+ \sqrt{2}\zeta\right]\\ \no 
 m_{\Xi^\ast}&=&g_D^S \left[(1+\alpha_{DS})\sigma+(2-\alpha_{DS})
\sqrt{2}\zeta\right]\\ \no 
 m_{\Omega}  &=& g_D^S \left[2 \alpha_{DS}\sigma+(3-\alpha_{DS})
\sqrt{2}\zeta\right]\no
\eea
The new parameters are connected to the parameters in
(\ref{spin32meson}) by $g_{D8}^W=-\sqrt{120} (1-\alpha_{DS}) g_D^S$ and 
$g_{D1}^W=\sqrt{90}  g_D^S$. $g_D^S$ and $\alpha_{DS}$ can now be fixed 
to reproduce the
masses of the baryon decuplet. 
As in the case of the nucleon, the 
coupling of the $\Delta$ to the strange condensate is nonzero.  

\label{bmver1}

It is desirable to have an alternative way of baryon mass generation, 
where the nucleon and the $\Delta$ mass depend only on $\sigma$. For the 
nucleon this can be accomplished for example by taking the limit 
$\alpha_{OS}=1$ and $g_{O1}^S=\sqrt{6}g_{O8}^S$. Then, 
the coupling constants between the baryon octet and the two scalar condensates 
are related to the additive quark model. This leaves only one coupling 
constant to adjust for the correct nucleon mass. For a fine-tuning 
of the remaining masses, it is necessary to introduce an explicit 
symmetry breaking term, that breaks the SU(3)-symmetry along the hypercharge 
direction. A possible term already discussed in \cite{paper2,sche69}, which 
respects the 
Gell-Mann-Okubo mass relation, is
\be 
\label{sche-esb}
  {\cal L}_{\Delta m} = -m_1 \Tr (\ovl{B} B - \ovl{B} B S)
  -m_2 \Tr (\ovl{B} S B) ,
\ee
where $S_{b}^a = -\frac{1}{3}[\sqrt{3} (\lambda_8)_{b}^a-\delta_{b}^a]$. 
As in the first case, only three coupling constants,  
$g_{N \sigma}\equiv 3 g_{O8}^S$, 
$m_1$ and $m_2$, are sufficient to reproduce the experimentally known 
baryon masses. Explicitly, the baryon masses have the values
\bea
\label{bmver2eq1}
m_N &=& -g_{N \sigma}  \sigma   \\ \no
m_{\Xi} &=&-\frac{1}{3} g_{N \sigma} \sigma -\frac{2}{3} g_{N \sigma} \sqrt{2}\zeta      +m_1  +m_2\\ \no
m_{\Lambda} &=& -\frac{2}{3} g_{N \sigma} \sigma-\frac{1}{3} g_{N \sigma} \sqrt{2}\zeta+\frac{m_1+2 m_2}{3} \\ \no
m_{\Sigma} &=& -\frac{2}{3} g_{N \sigma} \sigma -\frac{1}{3} g_{N \sigma} \sqrt{2}\zeta +m_1 , 
\eea   
For the baryon decuplet the choice $\alpha_{DS}=0$ yields
coupling constants related to the additive quark model. We introduce an explicit symmetry breaking
proportional to the number of strange quarks for a given baryon
species. Here we need only one additional parameter $m_{Ds}$ to obtain
the masses of the baryon decuplet:
\bea
\label{resmassen2}
 m_{\Delta} &=& g_{\Delta\sigma} \left[ 3 \sigma \right]\\ \no
 m_{\Sigma^\ast}&=& g_{\Delta\sigma} \left[ 2\sigma+ \sqrt{2}\zeta\right] + 
m_{Ds}\\ \no 
 m_{\Xi^\ast}&=&g_{\Delta\sigma} \left[ 1 \sigma+ 2\sqrt{2}\zeta\right] + 
2 m_{Ds}\\ \no 
 m_{\Omega}  &=& g_{\Delta\sigma} \left[0 \sigma+ 3 \sqrt{2}\zeta\right] + 
3 m_{Ds}\\ \no
\eea
For both versions of the baryon-meson interaction the parameters are fixed 
to yield the 
baryon masses of the octet and the decuplet. The corresponding
parameter set $C_2$, has been discussed in detail in \cite{paper3}.

\label{bmver2}
\subsection*{Vector mesons}
For the spin-$\frac 12$ baryons two independent interaction terms
with spin-1 mesons can be constructed, 
in analogy to the interaction of the baryon octet with the scalar mesons.
They correspond to the antisymmetric ($f$-type) and symmetric ($d$-type) 
couplings, 
respectively. 
From the universality 
principle \cite{saku69} and the vector meson dominance model one may 
conclude that the $d$-type 
coupling should be small. Here $\alpha_V=1$, 
i.e. pure $f$-type coupling, 
is used. It was shown in \cite{paper3}, that
a small admixture of d-type coupling 
allows for some fine-tuning of the single-particle energy levels of nucleons 
in nuclei. 
As in the case of scalar mesons, for 
$g_{O1}^V=\sqrt{6}g_{O8}^V$, the strange vector field $\phi_{\mu} \sim  
\ovl{s}\gamma_{\mu} s $  does not couple to the nucleon. The remaining 
couplings to the strange baryons are then determined by symmetry relations:
\parbox{7cm}
{\begin{eqnarray*}
          g_{N\omega} &=&(4 \alpha_V-1) g_{O8}^V \\ 
          g_{\Lambda \omega} &=& \frac{2}{3} (5 \alpha_V-2)g_{O8}^V\\   
          g_{\Sigma \omega} &=& 2 \alpha_V g_{O8}^V \\
          g_{\Xi \omega} &=& (2 \alpha_V-1) g_{O8}^V
\end{eqnarray*}} 
\hfill
\parbox{7cm}
{\begin{eqnarray*}          \\ 
          g_{\Lambda \phi} &=& - \frac{\sqrt{2}}{3}
                              (2\alpha_V+1)g_{O8}^V\\ 
          g_{\Sigma \phi} &=& - \sqrt{2} (2\alpha_V-1)g_{O8}^V \\  
          g_{\Xi \phi} &=& -2 \sqrt{2} \alpha_V g_{O8}^V \quad .           
\end{eqnarray*}}
\hfill
\parbox{1cm}
{\begin{eqnarray}
\end{eqnarray}}
\hfill
In the limit $\alpha_V=1$, the relative values of the coupling constants 
are related to the additive quark model via: 
\be
\label{quarkcoupling}
 g_{\Lambda \omega} = g_{\Sigma \omega} = 2 g_{\Xi \omega} = \frac{2}{3} 
 g_{N \omega}=2 g_{O8}^V \qquad 
 g_{\Lambda \phi} = g_{\Sigma  \phi} = \frac{g_{\Xi \phi}}{2} = 
   \frac{\sqrt{2}}{3} g_{N \omega}  .
\ee   
Note that all coupling constants are fixed once e.g. $g_{N\omega}$ 
is specified.
For the coupling of the baryon resonances to the vector mesons we
obtain the same Clebsch-Gordan coefficients as for the coupling to the
scalar mesons. This leads to the following relations between the
coupling constants:\\
\parbox{5cm}
{\begin{eqnarray*}
g_{\Delta\omega} &=& (3-\alpha_{DV}) g_{DV} \\ 
g_{\Sigma^\ast\omega}&=& 2 g_{DV} \\
g_{\Xi^\ast\omega}&=& (1+\alpha_{DV})  g_{DV}\\
g_{\Omega\omega}  &=& \alpha_{DV} g_{DV} 
\end{eqnarray*}}
\hfill
\parbox{6cm}
{\begin{eqnarray*}
g_{\Delta\phi} &=& \sqrt{2} \alpha_{DV} g_{DV} \\
g_{\Sigma^\ast\phi}&=& \sqrt{2} g_{DV} \\ 
g_{\Xi^\ast\phi}&=& \sqrt{2} (2-\alpha_{DV}) g_{DV} \\
g_{\Omega\phi}  &=&\sqrt{2}(3-\alpha_{DV})g_{DV} \quad . 
\end{eqnarray*}}
\hfill
\parbox{1cm}
{\begin{eqnarray}
\end{eqnarray}}
\hfill

In analogy to the octet case we set $\alpha_{DV}=0$, so that the strange 
vector meson $\phi$ does not couple to the $\Delta$-baryon.
The resulting coupling constants again obey the additive quark model
constraints:
\bea
\label{resveccoupling}
g_{\Delta \omega} &=& \frac{3}{2} g_{\Sigma^\ast \omega} = 
3 g_{\Xi^\ast \omega}=3 g_{DV} \qquad
g_{\Omega \omega} = 0 \\ \no
g_{\Omega \phi} &=&\frac32 g_{\Xi^\ast \phi}= 3  g_{\Sigma^\ast \phi}
= \sqrt{2} g_{\Delta \omega} \qquad g_{\Delta \phi} = 0 \no
\eea  
Hence all coupling constants of the baryon decuplet
are again fixed if one overall coupling $g_{DV}$ is specified. Since
there is no vacuum restriction on the $\Delta$-$\omega$ coupling, like
in the case of the scalar mesons, we have to consider different
constraints. This will be discussed in section \ref{hadmatter}.

\label{vector}
\section{Mean-field approximation}
The terms discussed so far involve the full quantum field operators. 
They cannot be treated exactly. Hence, to investigate hadronic matter
properties at finite baryon density  we 
adopt the mean-field approximation.
This nonperturbative relativistic 
method is applied to solve approximately the nuclear many body problem by replacing the 
quantum field operators by their classical expectation values (for a recent review 
see \cite{serot97}), i.e. the fluctuations around the vacuum 
expectation values of the field operators are neglected:
\bea
            \sigma(x)&=&\langle \sigma \rangle +\delta \sigma  
\rightarrow \langle \sigma \rangle \equiv \sigma \, ; \quad 
            \zeta(x)=\langle \zeta \rangle +\delta \zeta  
\rightarrow \langle \zeta \rangle \equiv \zeta \\ \no
      \omega_{\mu}(x) &=&\langle \omega \rangle \delta_{0 \mu}+ 
 \delta \omega_{\mu} 
\rightarrow  \langle \omega_0 \rangle \equiv \omega\, ; \quad 
      \phi_{\mu}(x) =\langle \phi \rangle \delta_{0 \mu}+ 
 \delta \phi_{\mu} 
\rightarrow  \langle \phi_0 \rangle \equiv \phi .
\eea
The fermions are treated as quantum mechanical single-particle operators. 
The derivative terms can be neglected and only the 
time-like component of the vector mesons 
$\omega \equiv \langle \omega_0 \rangle$ and 
$\phi \equiv \langle \phi_0 \rangle$ 
survive if we assume homogeneous and isotropic infinite baryonic
 matter. Additionally, due to 
 parity conservation we have $\langle \pi_i \rangle=0$.
The baryon resonances are treated as spin-$\frac 12$ particles with
spin-$\frac 32$ degeneracy.
After these approximations the Lagrangian (\ref{lagrange}) 
reads
\begin{eqnarray*}
{\cal L}_{BM}+{\cal L}_{BV} &=& -\sum_{i} \overline{\psi_{i}}[g_{i 
\omega}\gamma_0 \omega^0 
+g_{i \phi}\gamma_0 \phi^0 +m_i^{\ast} ]\psi_{i} \\ \no
{\cal L}_{vec} &=& \frac{ 1 }{ 2 } m_{\omega}^{2}\frac{\chi^2}{\chi_0^2}\omega^
2  
 + \frac{ 1 }{ 2 }  m_{\phi}^{2}\frac{\chi^2}{\chi_0^2} \phi^2
+ g_4^4 (\omega^4 + 2 \phi^4)\\
{\cal V}_0 &=& \frac{ 1 }{ 2 } k_0 \chi^2 (\sigma^2+\zeta^2) 
- k_1 (\sigma^2+\zeta^2)^2 
     - k_2 ( \frac{ \sigma^4}{ 2 } + \zeta^4) 
     - k_3 \chi \sigma^2 \zeta \\ 
&+& k_4 \chi^4 + \frac{1}{4}\chi^4 \ln \frac{ \chi^4 }{ \chi_0^4}
 -\frac{\delta}{3}\ln \frac{\sigma^2\zeta}{\sigma_0^2 \zeta_0} \\ \no
{\cal V}_{SB} &=& \left(\frac{\chi}{\chi_0}\right)^{2}\left[m_{\pi}^2 f_{\pi} 
\sigma 
+ (\sqrt{2}m_K^2 f_K - \frac{ 1 }{ \sqrt{2} } m_{\pi}^2 f_{\pi})\zeta 
\right] , 
\end{eqnarray*}
with the effective mass $m_i^\ast$ of the baryon $i$, which 
is defined according to section \ref{bm} for  
$i=N,\Lambda,\Sigma,\Xi,\Delta,\Sigma^\ast,\Xi^\ast,\Omega$.\\
Now it is straightforward to write down the expression 
for the thermodynamical potential of the grand canonical 
ensemble, $\Omega$, per volume $V$ 
at a given chemical potential $\mu$ and at zero temperature:
\be
   \frac{\Omega}{V}= -{\cal L}_{vec} - {\cal L}_0 - {\cal L}_{SB}
-{\cal V}_{vac}- \sum_i \frac{\gamma_i }{(2 \pi)^3}  
\int d^3k [E^{\ast}_i(k)-\mu^{\ast}_i]   
\ee 
The vacuum energy ${\cal V}_{vac}$ (the potential at $\rho=0$) 
has been subtracted in 
order to get a vanishing vacuum energy. The $\gamma_i$ denote the fermionic 
spin-isospin degeneracy factors.
The single particle energies are 
$E^{\ast}_i (k) = \sqrt{ k_i^2+{m_i^*}^2}$ 
and the effective chemical potentials read
 $\mu^{\ast}_i = \mu_i-g_{\omega i} \omega-g_{\phi i} \phi$.\\ 
The mesonic fields are determined by extremizing $\frac{\Omega}{V}(\mu, T=0)$:
\bea
\label{tbgls}
\frac{\partial (\Omega/V)}{\partial \chi} &=& 
        -\omega^2 m_{\omega}^2 \frac{\chi}{\chi_0^2} 
        + k_0 \chi (\sigma^2+\zeta^2) 
        - k_3 \sigma^2 \zeta 
        + \left( 4 k_4 + 1 + 4 \ln \frac{ \chi }{\chi_0}
        - 4 \frac{\delta}{3} \ln \frac{\sigma^2 \zeta}{\sigma_0^2\zeta_0}
        \right) \chi^3  +\\ \no
&+&2\frac{\chi}{\chi_0^2}\left[m_{\pi}^2 f_{\pi}\sigma +(\sqrt{2}m_K^2 f_K - 
\frac{ 1 }{ \sqrt{2} }
 m_{\pi}^2 f_{\pi}) \zeta \right] =0\\    
\frac{\partial (\Omega/V)}{\partial \sigma} &=& 
 k_0 \chi^2 \sigma - 4 k_1 (\sigma^2+\zeta^2)\sigma 
 - 2k_2 \sigma^3        - 2 k_3 \chi \sigma \zeta  
  -2\frac{\delta \chi^4}{3 \sigma} +\\ \no
&+& \left(\frac{\chi}{\chi_0}\right)^{2} m_{\pi}^2 f_{\pi}  
+ \sum_{i} \frac{\pa m_i^{\ast}}{\pa \sigma}\rho^s_{i}=0 \\        
\frac{\partial (\Omega/V)}{\partial \zeta} &=&  
   k_0 \chi^2 \zeta - 4 k_1 (\sigma^2+\zeta^2) \zeta 
- 4 k_2 \zeta ^3 - k_3 \chi \sigma^2
  -\frac{\delta \chi^4}{3 \zeta} +\\ \no
 &+& \left(\frac{\chi}{\chi_0}\right)^{2}  \left[\sqrt{2}m_K^2 f_K 
- \frac{ 1 }{ \sqrt{2}} m_{\pi}^2 f_{\pi}\right]+\sum_{i} 
\frac{\pa m_i^{\ast}}{\pa \zeta} \rho_i^s=0 \\ 
\frac{\partial (\Omega/V)}{\partial \omega} &=& -\left(\frac\chi\chi_0\right)
m_{\omega}^2 \omega - 4 g_4^4 \omega^3 + \sum_i \frac{g_{i \omega }}
{\rho_i} = 0 \\ 
\frac{\partial (\Omega/V)}{\partial \phi} &=& -\left(\frac\chi\chi_0\right)
m_{\phi}^2 \phi - 8 g_4^4 \phi^3 + \sum_i \frac{g_{i\phi}}{\rho_i} = 0 
\eea
The scalar densities  $\rho^s_{i}$ and the vector densities $\rho_i$
can be calculated analytically for the case $T=0$, yielding
\bea
\rho^s_i &=& \gamma_i
\int \frac{d^3 k}{(2 \pi)^3} \frac{m_i^{\ast}}{E^{\ast}_i} = 
\frac{\gamma_i  m_i^{\ast}}{4 \pi^2}\left[ k_{F i} E_{F i}^{\ast}-m_i^{\ast 2} 
\ln\left(\frac{k_{F i}+E_{F i}^{\ast}}{m_i^{\ast}}\right)\right] \\ 
 \rho_i &=&  \gamma_i \int_0^{k_{F i}} \frac{d^3 k}{(2 \pi)^3} =
\frac{\gamma_i k_{F i}^3}{6 \pi^2}        \,   .
\eea
The energy density and the pressure  follow from the Gibbs--Duhem relation, 
$\epsilon = \Omega/V+ \sum_i \mu_i \rho^i$ and $p= - \Omega/V$. 
The Hugenholtz--van Hove theorem \cite{hugo58} yields the Fermi surfaces 
as $ E^{\ast}(k_{F i})= \sqrt{k_{F i}^2+m_i^{\ast 2}} 
= \mu^{\ast}_i $ .

\label{mfa}
%
\section{Results for dense nuclear matter}
\label{hadmatter}
\subsection{Parameters}
Fixing of the parameters to vacuum and nuclear
matter ground state properties was discussed in detail in \cite{paper3}.
It has
been shown that the obtained parameter sets describe the nuclear
matter saturation point, hadronic vacuum masses and properties of
finite nuclei reasonably well. The additional
parameters here are the couplings of the baryon resonances to the
scalar and vector mesons. For the scalar mesons this is done by a fit
to the vacuum masses of the spin-$\frac 32$ baryons.
The coupling of the baryon resonances to the spin-1 
mesons will be discussed later.  These new parameters
will not influence the results for normal nuclear matter and finite nuclei.

\subsection {Extrapolation to high densities}
Once the parameters have been fixed to nuclear matter at $\rho_0$, the
condensates and hadron masses at high baryon densities can be
investigated, assuming that the change of the parameters of the
effective theory with density are small.
The behaviour of the fields and the masses of the baryon
octet have been investigated in \cite{paper3}.
It is found that the gluon condensate 
$\chi$ stays nearly constant when the density increases. This implies that
the approximation of a frozen glueball is reasonable.
In these calculations the strange condensate $\zeta$ is 
only reduced by about 10 percent from
its vacuum expectation value. This is not surprising since there are only 
nucleons in the system and the nucleon--$\zeta$ coupling is fairly weak.
The main effect occurs for the non--strange condensate $\sigma$: This 
field drops to 30 percent of its vacuum expectation value at 
4 times normal nuclear density, at even higher densities
the $\sigma$ field saturates.
The behaviour of the condensates is also reflected in the behaviour of
the baryon masses: The
change of the scalar fields causes a change of the
baryon masses in the dense medium.
Furthermore, the change of the baryon masses depends on 
the strange quark content of the baryon. This is due to
the different coupling of the baryons to the 
non-strange and strange condensate.
The masses of the vector mesons are shown in fig. \ref{vmassenrho}. 
The corresponding terms in the lagrangean are
discussed in \cite{paper3}. These masses stay nearly 
constant when the density is increased.

\begin{figure}
\centerline{\psfig{figure=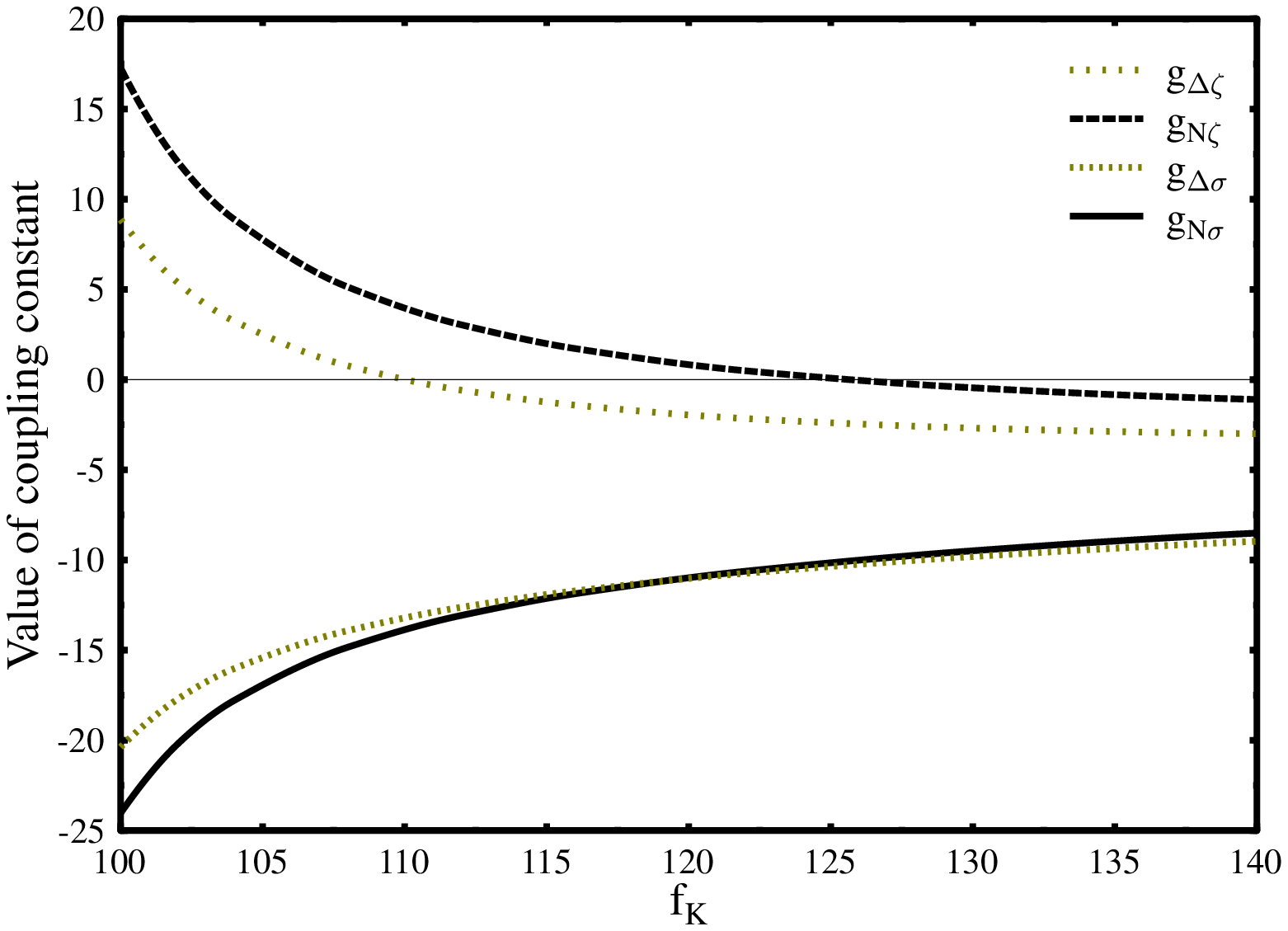,width=18cm,height=18cm}}
\caption{\label{zetacoup} Coupling of the nucleon and the $\Delta$ to the
non-strange ($\sigma$) and strange ($\zeta$) scalar condensates as a
function of the kaon decay constant $f_K$. For 
$f_K \approx 115-125 MeV$ the coupling of the nucleon and the $\Delta$
 to the non-strange scalar 
field $\sigma$ are nearly equal. The coupling strength to the strange
scalar field are different in sign. This results in
the mass difference of $\Delta m \approx 300$MeV in the vacuum.}
\end{figure}

\begin{figure}
\hspace{-2cm}
\centerline{\psfig{figure=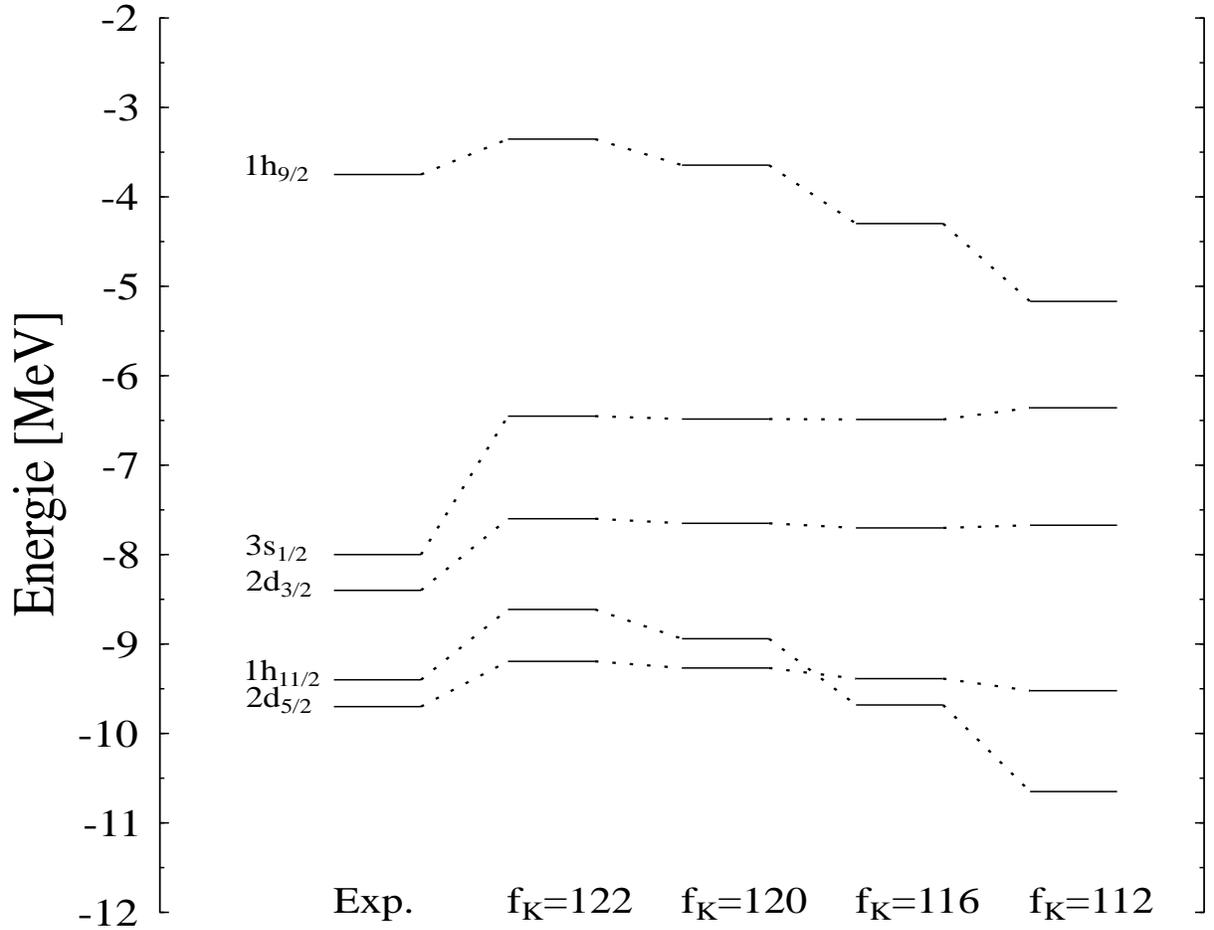,width=14cm,height=12cm}}
\vspace{1.5cm}
\caption{\label{kernefk} Single particle energie levels for protons
in $Pb^{208}$ for parameter sets with
different values of the kaon decay constant $f_k$ (in $MeV$). The
smaller the decay constant is chosen, the more the gap between the
levels $1h_{\frac 92}$ and $3s_{\frac 12}$ decreases. The parameter sets
were obtained by just variing $f_k$ in $C_1$.}  
\end{figure}

\begin{figure}
\epsfig{figure=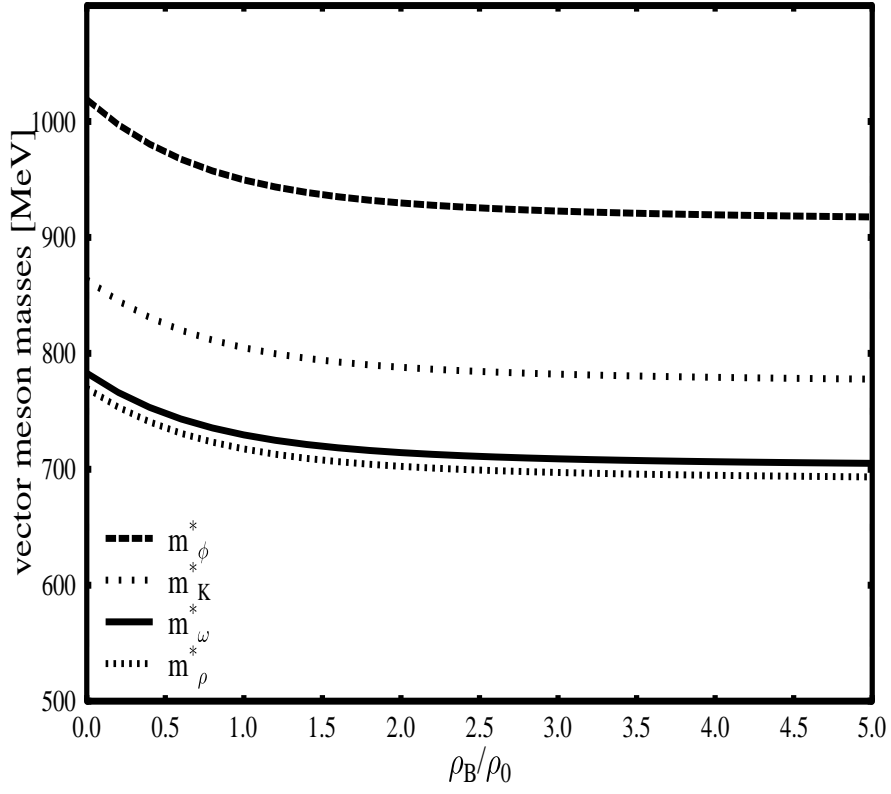,width=14cm,height=14cm}
\caption{\label{vmassenrho}Vector meson masses  as a function
of density.}
\end{figure}

\begin{figure}
\epsfig{figure=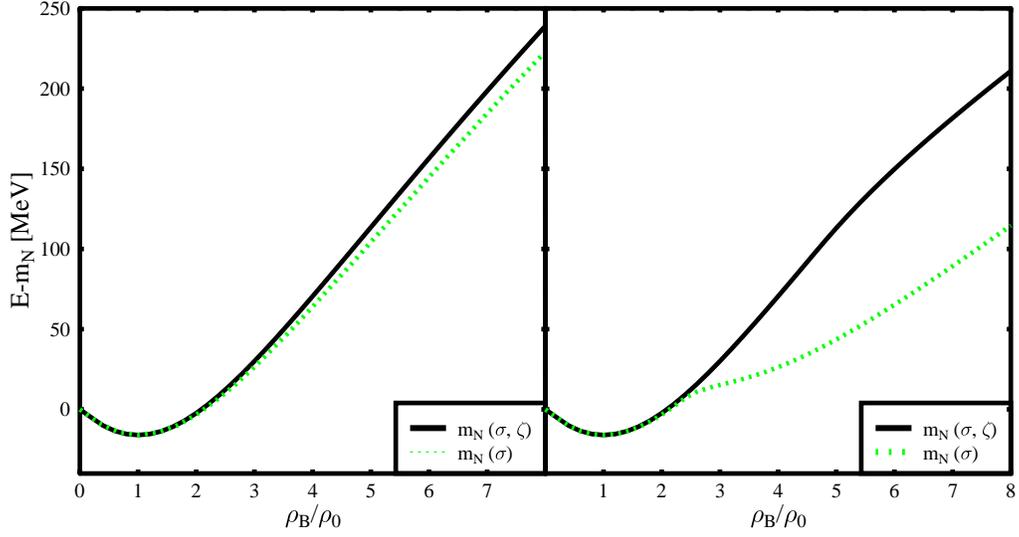,width=16cm}
\caption{\label{eos} Equation of state for infinite nuclear matter for
the parameter sets $C_1$ ($m_N=m_N(\sigma,\zeta)$) and $C_2$ 
($m_N=m_N(\sigma)$). In the left picture resonances are
neglected while they are included in the right picture. If the
strange condensates couples to the nucleon the
influence of the $\Delta$ resonances on the equation of state is much weaker.}
\end{figure}

\label{baryonres}
Now we discuss the inclusion of baryonic spin-$\frac 32$ 
resonances. How do they affect the behaviour of dense hadronic matter?
We consider the 
two parameter sets $C_1$ and $C_2$, which satisfactorily describe finite 
nuclei \cite{paper3}. As stated
above, the main difference between the two parameter sets is the 
coupling of the strange condensate to the nucleon and to the $\Delta$. 
In $C_2$ this coupling is set to zero, while 
the nucleon and the $\Delta$ couple to
the $\zeta$ field in the case of $C_1$. 
Fig. \ref{zetacoup} shows how the strength of 
the coupling of the strange condensate
to the nucleon and the $\Delta$ depends on the vacuum 
expectation value of the strange
condensate $\zeta_0$. $\zeta_0$ in turn is a function of the kaon decay
constant ($\zeta_0 = \frac 1{\sqrt{2}} (f_\pi-f_K)$). The results are
obtained by changing the value of $f_K$, starting from parameter
set $C_1$. $f_K$ is expected to be in the range of 105 to 125
MeV \cite{qcdsum}. 
For infinite nuclear matter one obtains good fits for 
the whole range of expected values. 
But when these parameter sets are used to describe finite nuclei, 
satisfactory results are only obtained for a small range of values for
$f_K$, as can be seen for the proton single particle levels in 
fig. \ref{kernefk}: with decreasing $f_K$ the gap between the
single-particle levels $1h_{\frac 92}$ and $3s_{\frac 12}$ in
$^{208}Pb$ decreases such that
e.g. for $f_K=112 MeV$ the experimentally observed shell closure
cannot be reproduced in the calculation. This result is not very surprising,
because the smaller value of $f_K$ leads to a stronger coupling of the
nucleon to the strange field, with a mass of $m_\zeta \approx 1
GeV$. But it has been shown \cite{tang95}, that for a reasonable 
description of finite nuclei the nucleon must mainly couple to a
scalar field with $m \approx 500-600 MeV$.
The equation of state of dense hadronic matter for vanishing
strangeness is shown in Fig. \ref{eos}.
Here two $C_1$ fits are compared, one with  $f_k=122$, which 
corresponds to the fit that has been tested to describe finite nuclei
satisfactory in \cite{paper3}, and a $C_1$-type fit with
$f_k= 116$, as the minimum acceptable value extracted from Fig.
\ref{kernefk}. The resulting values of coupling-constants
to the nucleon are $g_{N\zeta}\approx 0.49$ for $f_K=122$ MeV and 
 $g_{N\zeta}\approx 1.72$ for $f_K=116$ MeV. For the $\Delta$-baryon 
 $g_{\Delta\zeta}\approx -2.2$ and $g_{\Delta\zeta} \approx -0.59$, 
respectively. If these values are compared to the couplings to the
non-strange condensate (which is around $-10$ for the nucleon and the
$\Delta$ in both cases) one observes that the mass difference between
nucleon and $\Delta$ is due to the different coupling to the strange
condensate.

\begin{figure}
\epsfig{figure=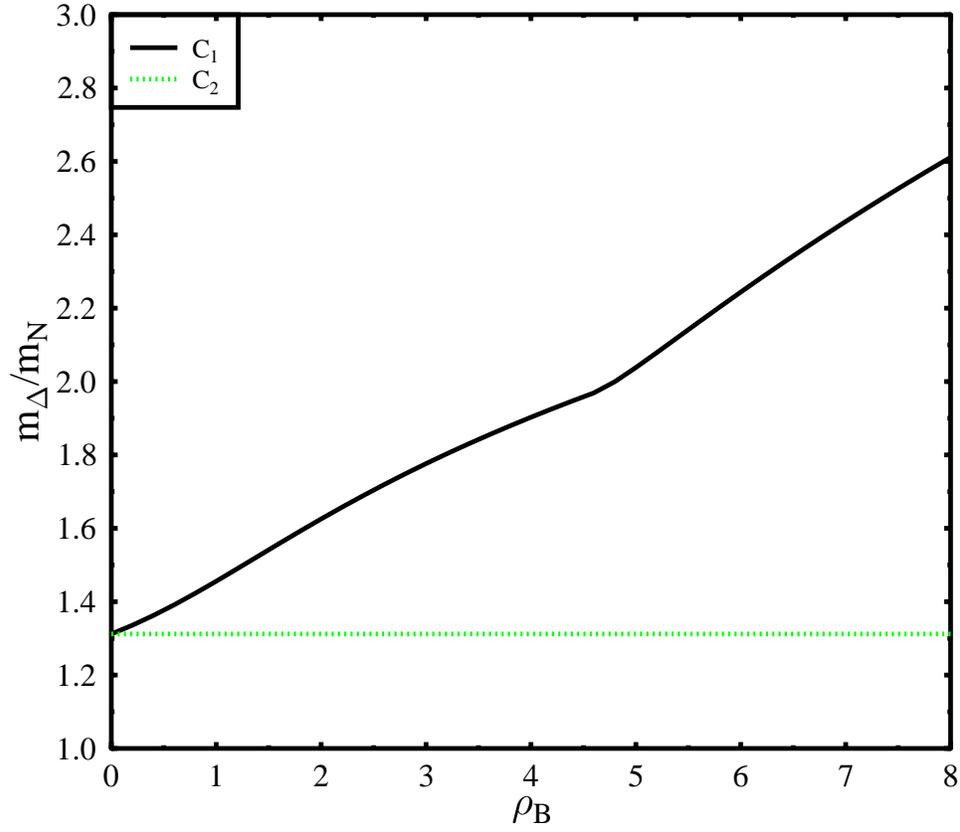,width=14cm}
\caption{\label{massratio} Ratio of $\Delta$-mass to nucleon-mass as a
function of density for the two parameter sets $C_1$ and $C_2$.}
\end{figure}

\begin{figure}
\begin{center}
\epsfig{figure=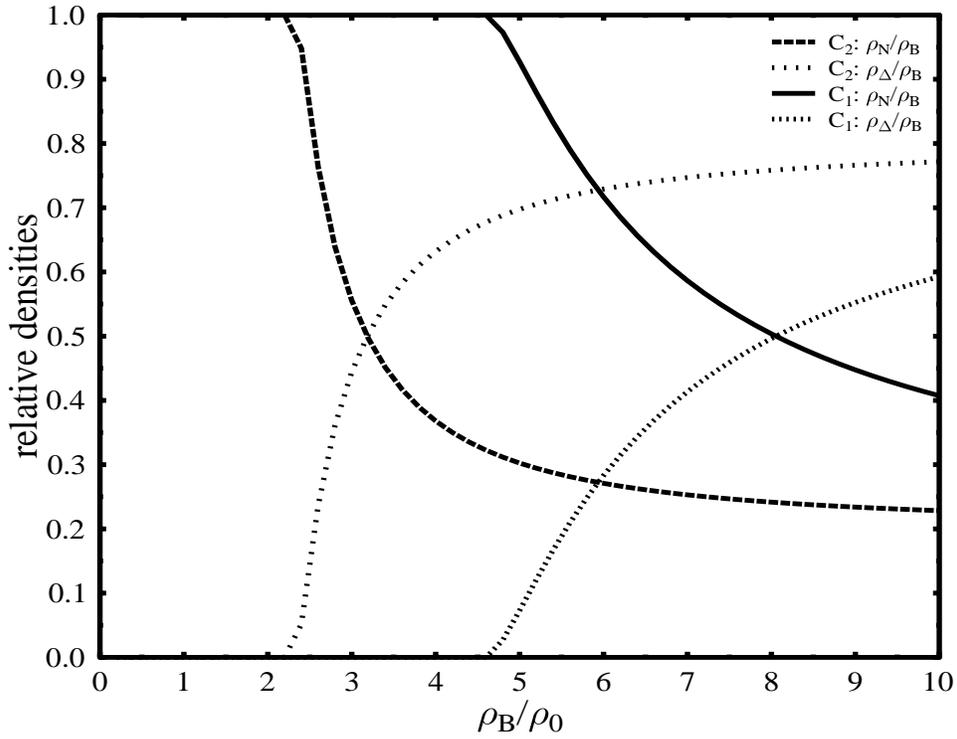,width=16cm,height=20cm}
\end{center} 
\caption{\label{relden} Relative densities of nucleons and $\Delta$`s for
various parameter sets. The production rate of $\Delta$'s depends strongly 
on the parameter set, i.e. on the strength of the nucleon-$\zeta$ and 
$\Delta$-$\zeta$ coupling.}
\end{figure}

Furthermore the resulting equation of state for parameter set $C_2$
is plotted. Here the nucleon and $\Delta$-mass do not depend on the 
strange condensate. Fig. \ref{eos} shows
two main results:
The resulting EOS does not change 
significantly if $f_K$ in the $C_1$-fits is varied within the reasonable
range discussed above. In the following we refer to the $C_1$-fit of 
\cite{paper3}
with $f_K=122 MeV$. \\
However, the different ways of 
nucleon and $\Delta$ mass generation lead to drastic
differences in the resulting equations of state:
 
A pure $\sigma$-dependence of the masses of the nonstrange baryons
($C_2$) leads to an 
equation of state which is strongly influenced by the production of resonances
at high densities. This is not the case 
when both masses are partially generated by the strange condensate
($C_1$), Fig. \ref{eos}.
In both fits the coupling of the $\Delta$ to the $\omega$-meson 
($g_{\Delta\omega}$) has been set equal to $g_{N\omega}$. The very
different behaviour of the EOS can be understood from
the ratio of the effective $\Delta$-mass to the
effective nucleon-mass, Fig. \ref{massratio}. If the coupling of the
nucleon to the $\zeta$ field is set to zero ($C_2$), 
the mass ratio stays at the constant value 
$\frac {m_\Delta}{m_N} =\frac {g_{\Delta\sigma}}{g_{N\sigma}}\approx 1.31$.
However, if the nucleon couples to the strange condensate ($C_1$), the mass
ratio $\frac {m_\Delta}{m_N}$ increases with density, due to the
different coupling of the nucleon and the $\Delta$
to the strange condensate $\zeta$. The
$\Delta$ does not feel less scalar attraction - the coupling to the
$\sigma$ field is the same for the nonstrange baryons. However, the mass
of the $\Delta$ does not drop as fast as in the case of pure
$\sigma$-couplings, and hence the production of baryon resonances is less
favorable at high densities, Fig. \ref{relden}. 

Both coupling constants of the $\Delta$-baryon are
freely adjustable in the RMF models \cite{sero86,bogu83,koso98}. 
In the chiral model, which incorporates 
dynamical mass generation, the scalar couplings are fixed by the
corresponding vacuum masses. If explicit symmetry
breaking for the baryon mass generation is neglected, then 
the scalar couplings are fixed by the vacuum alone.
To investigate the influence of the coupling to the
strange condensate $\zeta$, small explicitly symmetry
breaking terms $m1,m2$ are used. This model behaves similar as the
RMF models with 
$r=\frac {g_{\Delta\sigma}}{g_{N\sigma}}=\frac {m_\Delta}{m_N}$.

The remaining problem is the coupling of the resonances to the
vector mesons. The coupling constants can be restricted 
by the requirement that resonances are absent 
in the ground state of normal nuclear matter. Furthermore possible 
secondary minimua in the nuclear equation of state should lie above the
saturation energy of normal nuclear matter. 

QCD sum-rule
calculations suggest \cite{qcdsum} that the net attraction 
for $\Delta$`s in nuclear
matter is larger than that of the nucleon. From these constraints a
'window' of possible parameter sets $g_{\Delta\sigma},
g_{\Delta \omega}$ has been extracted \cite{koso98}. 
In the chiral model one then obtains for each type of mass generation
 only a small region of possible
values for $g_{\Delta\omega}$.
The $\Delta-\omega$ coupling in Fig.7 is in this range. 
Pure $\sigma$-coupling ($C_2$) of the
non-strange baryons yields a range of coupling constants 
$r_v=\frac {g_{\Delta\omega}}{g_N\omega}$ between 
${0.91 < r_v < 1}$. For a non-vanishing
$\zeta$-coupling one obtains ${0.68 < r_v < 1}$.
A smaller value of the ratio $r_v$ 
(less repulsion), leads to
higher $\Delta$-probabilities and to softer equations of state. 
Due to this freedom in the coupling of the resonances to the vector mesons
the equation of state cannot be predicted unambigiously from the chiral model.
Here additional input from experiments are necessary to
pin down the equation of state. \\
Finally we address the question, whether at very high densities the
anti-nucleon potentials become overcritical. That means the potential 
for anti-nucleons may become larger than 2 $m_N c²$ and nucleon-
anti-nucleon pairs may be spontaneously emitted \cite{mis90}. 
The nucleon and anti-nucleon
potentials in the chiral model are shown as function of density 
(Fig. \ref{nucpot}) for parameter
set $C_1$ with and without quartic vector self-interaction.
The latter is to obtain reasonable compressibility in the chiral model 
\cite{paper3} and is in agreement with the
principle of naturalness stated in \cite{tang95}. From that the 
anti-nucleon potentials are predicted not to turn overcritical 
at densities below $12 \rho_0$ in the chiral model (Fig. \ref{nucpot} left).
Earlier calculations in RMF-models \cite{mis90} did not include the 
higher order vector self-interactions. Then
spontaneous anti-nucleon production occurs around $4-6 \rho_0$. This also
happens in the chiral model if the quartic-terms would be
neglected (Fig. \ref{nucpot} right). 
The critical density shifts to even 
higher values, if the equation of state is
softened by the baryon resonances, as can be seen in Fig. \ref{nucpotres}.
Hence, the chiral mean field model does not predict
overcriticality for reasonable densities.

%
\section{conclusion}
\label{conclusion}
Spin-$\frac 32$-baryon resonances can be included
consistently in the nonlinear chiral SU(3)-model.
The coupling constants of the baryon resonances to the
scalar mesons are fixed by the vacuum masses. Two different ways of 
mass generation were investigated. It is found that
they lead to very different predictions for the resulting equation of
state of non-strange nuclear matter.
The coupling of the baryon resonances to the vector mesons cannot be
fixed. The allowed range of this coupling constant is
restricted by requireing that possible density isomers are not
absolutely stable, that there are no $\Delta$'s in the nuclear matter
ground state and by QCD sum-rule induced assumption that the net
attraction of $\Delta$'s in nuclear matter is larger than that for
nucleons. Nevertheless, the behaviour of non-strange nuclear matter
cannot be predicted unambigiously within the chiral $SU(3)$-model, so that
further experimental input on $\Delta$-production in high density
systems  and
theoretical investigations on how the resonance
production influences the observables in these systems 
(neutron stars, heavy ion-collisions) is needed. 
For both cases calculations are under way \cite{hanau99,hanau00,ziezu}.

\begin{acknowledgements}
The authors are grateful to C. Beckmann, L. Gerland, I. Mishustin,
L. Neise, and S. Pal 
for fruitful discussions. This work is supported by Deutsche 
Forschungsgemeinschaft (DFG), Gesellschaft f\"ur Schwerionenforschung
(GSI), Bundesministerium f\"ur Bildung und Forschung (BMBF) and
Graduiertenkolleg Theoretische und experimentelle Schwerionenphysik.
\end{acknowledgements}
\appendix
\section{}
\label{append}
The SU(3) matrices of the hadrons are (suppressing the Lorentz indices)
\begin{displaymath}
X=\frac{1}{\sqrt{2}}\sigma^a \lambda_a=
\left( 
\begin{array}{ccc}
   (a_0^0  +\sigma)/\sqrt{2} & a_0^+ & \kappa^+\\   
    a_0^- & (-a_0^0+\sigma)\sqrt{2} & \kappa^0 \\
   \kappa^- & \ovl{\kappa^0}& \zeta 
\end{array} 
\right)
\end{displaymath}
\be
\label{psmatrix}
P =\frac{1}{\sqrt{2}}\pi_a \lambda^a
=\left ( \begin{array}{ccc} 
  \frac{1}{\sqrt{2}}\left ( \pi^0+{\frac {\eta^8}{\sqrt {1+2\,{w}^{2}}}}\right )&\pi^{+}
 &2\,{\frac {K^{+}}{w+1}}\\
  \noalign{\medskip}\pi^{-}&\frac{1}{\sqrt{2}}\left 
 (- \pi^0+
 {\frac {\eta^8}{\sqrt {1+2\,{w}^{2}}}}\right )&2\,{\frac { K^0
  }{w+1}}\\\noalign{\medskip}2\,{\frac {K^-}{w+1}}&2\,{\frac { 
 \ovl{K}^0}{w+1}}&-{\frac {\eta^8\,\sqrt {2}}{\sqrt {1+2\,{w}^{2}}}}
 \end {array}\right ) 
\ee
\be
\label{vmatrix}
V=\frac{1}{\sqrt{2}}v^a \lambda_a=
\left( \begin{array}{ccc}
   (\rho_0^0  +\omega)/\sqrt{2} & \rho_0^+ & K^{\ast +}\\   
    \rho_0^- & (-\rho_0^0+\omega)/\sqrt{2} & K^{\ast 0} \\
   K^{\ast -} & \ovl{K^{\ast 0}}& \phi
\end{array} \right)
\ee
\be
\label{bmatrix}
B=\frac{1}{\sqrt{2}}b^a \lambda_a=
\left( \begin{array}{ccc}
   \frac{\Sigma^0}{\sqrt{2}} +\frac{\Lambda^0}{\sqrt{6}}& \Sigma^+ & p\\   
    \Sigma^- & -\frac{\Sigma^0}{\sqrt{2}} +\frac{\Lambda^0}{\sqrt{6}} & n \\
   \Xi^- & \Xi^0& -2 \frac{\Lambda^0}{\sqrt{6}}
\end{array} \right)
\ee
for the scalar ($X$), pseudoscalar($P$), vector ($V$), baryon ($B$) 
and similarly for the axial vector meson fields.
A pseudoscalar chiral singlet
$Y=\sqrt{2/3} \eta_0  \, 1{\hspace{-2.5mm}1}$ can be added separately, since 
only an octet is allowed to enter the exponential.

The notation follows the convention of the Particle Data Group
(PDG),\cite{pdg98}, though we are aware of the difficulties to directly 
identify the scalar mesons with the physical particles \cite{sche98}. However, 
note that there is increasing evidence that supports the existence of a low-mass, 
broad scalar resonance, the $\sigma(560)$-meson, as well as a light strange scalar 
meson, the $\kappa(900)$ (see \cite{black98} and references therein).

The masses of the various hadrons are generated through their couplings to 
the scalar condensates, which are produced via spontaneous symmetry
breaking in the sector of the scalar fields. 
Of the 9 scalar mesons in the matrix $X$ only the 
vacuum expectation values of the 
components proportional to $\lambda_0$ and to the 
hypercharge $Y \sim \lambda_8$ are non-vanishing, and the vacuum expectation 
value $\langle X \rangle$ reduces to: 
\be
\label{vev}
\langle X \rangle=\frac {1}{\sqrt{2}}(\sigma^0 \lambda_0+\sigma^8 \lambda_8)
\equiv \mbox{diag } (\frac{\sigma}{\sqrt{2}} \,, \frac{\sigma}{\sqrt{2}} \,, 
\zeta ) , 
\ee
in order to preserve parity invariance and 
assuming, for simplicity, $SU(2)$ symmetry\footnote{This implies that 
isospin breaking effects will not occur, i.e., all hadrons of the 
same isospin multiplet will have identical masses. The electromagnetic 
mass breaking is neglected.} of the vacuum. 

\bibliography{chiral}
\bibliographystyle{prsty}
\clearpage

\begin{figure}
\begin{center}
\epsfig{figure=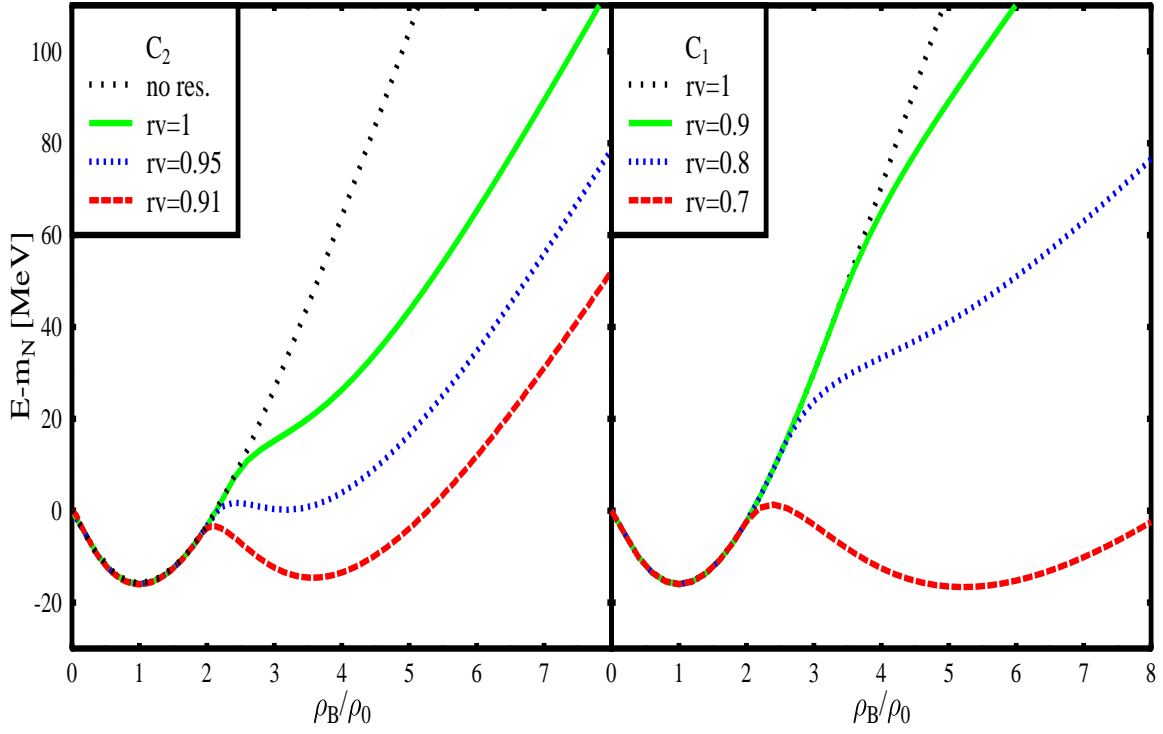,width=16cm,height=14cm}
\end{center}
\caption{\label{energierv} Equation of state for parameter sets
$C_1$ and $C_2$ for different values of the quotient 
$r_v=\frac{g_{N\omega}}{g_{\Delta\omega}}$. For the $C_2$-fit the value of
$r_v$ should not be less than $0.91$ to avoid the density isomer
being absolutely stable. For $C_2$, $r_v$ must be larger than
$0.68$. }
\end{figure}
\begin{figure}
\begin{center}
\epsfig{figure=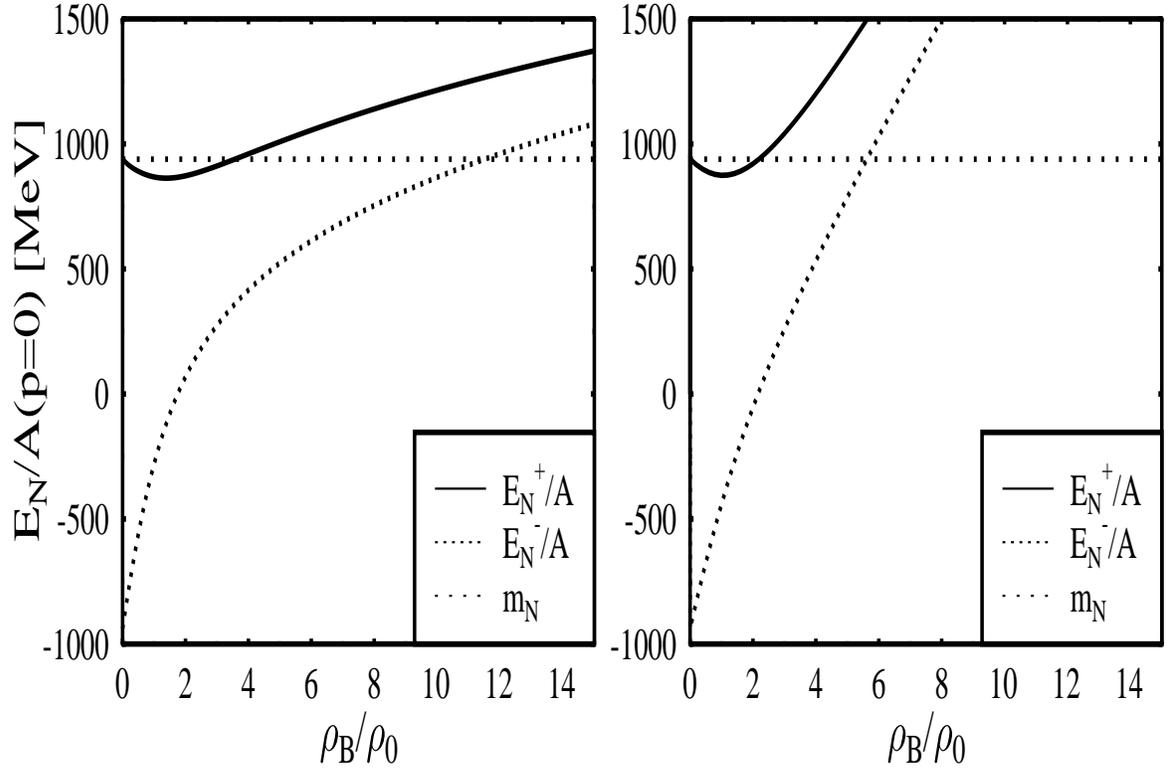, width=18cm, height=12cm}
\end{center}
\caption{\label{nucpot} Nucleon and anti-nucleon energy at $\vec{p}=0$
as a function of baryon density. On the left hand side parameter set
$C_1$ was use, while on the right hand side the coupling constant g4
for the quartic vector meson interaction was set to zero.}
\end{figure} 
\begin{figure}
\newpage
\epsfig{figure=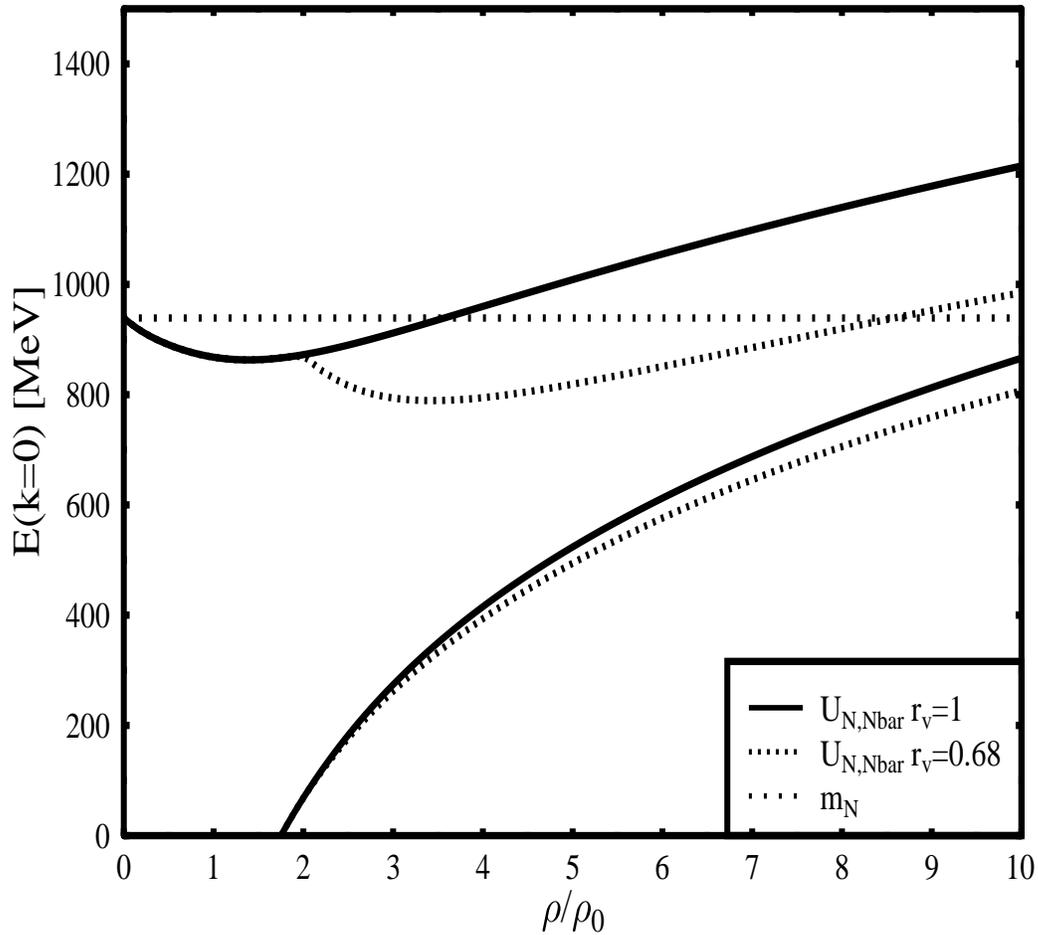, width=16cm, height=16cm}
\caption{\label{nucpotres} Nucleon and anti-nucleon energy at $\vec{p}=0$
as a function of baryon density using paramter set $C_1$ with $r_v=1$
and $r_v=0.68$. Lower values of the $\Delta-\omega$-coupling
lead to a significant change of the equation of state
(see fig.\ref{energierv}) and to an increase of the critical density
to even higher values.}
\end{figure} 
\end{document}